# Magnetosphere of a Spinning String

G. Chapline[1] and J. Barbieri[2]★

In this note we observe that the exact Maxwell-Einstein equations in the background metric of a spinning string can be solved analytically. This allows us to construct an analytical model for the magnetosphere which is approximately force free near to the spinning string. As in the case of a Kerr black hole in the presence of an external magnetic field the spinning string will acquire an electric charge which depends on the vorticity carried by the spinning string. The self-generated magnetic field and currents strongly resemble the current and magnetic field structure of the jets associated with active galaxies as they emerge from the galactic center.

The physical origin of the well-collimated jets seen emerging from the massive compact objects found at the centers of many galaxies has long been a mystery. Although the general idea [1] that the energy to power these jets comes from the electromagnetic torque exerted on the compact object by an accretion disk has long been accepted, it has never been demonstrated in detail how this might work. One problem is that in the case of a Kerr black hole model for the compact object the relevant equations are sufficiently nonlinear and intractable that one must rely on numerical methods. This complexity has so far prevented computational astrophysicists from either proving or disproving that externally generated magnetic fields interacting with a Kerr black hole can produce the observed jet structures. In this paper we suggest an alternative approach. We observe that the exact Maxwell-Einstein equations for the case of a spinning string surrounded by a static cylindrically symmetric configuration of charge and current densities can be solved analytically. This result allows us to find an analytical model for the magnetosphere that is very nearly force-free near to the spinning string, and confining farther away. Remarkably the structure of the self-generated magnetic field and currents in this model strongly resemble that inferred from mm-wave VLBLI observations of the astrophysical jets associated with the massive compact objects at the centers of active galaxies [2]. The success of this simple model suggests that the powerful jets seen emerging from the nuclei of active galaxies actually originate inside the compact object itself.

In 1987 P. Mazur discovered a remarkable solution to the Einstein equations in which space-time is flat everywhere except for a torsion-like singularity along a straight line where general relativity breaks down [3]. If we identify this straight line with the z-axis, the metric for the external space-time in cylindrical coordinates is

$$ds^2 = (cdt + Ad\phi)^2 - dr^2 - r^2 d\phi^2 - dz^2. \qquad (1)$$

where $A$ is a constant defining the vorticity of the spinning string ($A$ plays a role for the spinning string space-time similar to the angular momentum per unit mass of a black hole). Electromagnetic fields in this space-time background satisfy the covariant Maxwell equations:

$$\frac{1}{\sqrt{-g}}\frac{\partial}{\partial x^\beta}\left[\sqrt{-g}g^{\alpha\mu}g^{\beta\nu}F_{\mu\nu}\right]=-\mu_0 J^\alpha \tag{2}$$

where $J^\alpha = \rho dx^\alpha/dt$, $g^{00} = (1-A^2/r^2)/c^2$, $g^{0\phi} = A/cr^2$, $g^{\phi\phi} = -1/r^2$, $g^{rr} = g^{zz} = -1$, and $-g = c^2r^2$. If the charge and current densities depend only on r, the equations for the physical currents $j^\alpha = \sqrt{|g_{\alpha\alpha}|}J^\alpha$ become

$$j^r = 0$$

$$\mu_0 j^\phi = -\frac{\partial}{\partial r}\left(B^z - \frac{A}{cr}E^r\right)$$

$$\mu_0 j^z = \frac{1}{r}\frac{\partial}{\partial r}\left[rB^\phi\right] \tag{3}$$

$$\frac{\rho}{\varepsilon_0} = \frac{1}{r}\frac{\partial}{\partial r}\left[r(1-\frac{A^2}{r^2})E^r - AcB^z\right]$$

An exact analytic solution for these equations can be found by noting that combining the equations for $j_\phi$ and $\rho$ yields first order linear differential equations for $E_r$ and $B_z$ that can be solved by elementary methods. In particular, Eq's. (3) imply that for a stationary configuration of charge and current densities surrounding the spinning string the exterior electric and magnetic fields have the form

$$E_r = \frac{1}{\varepsilon_0}\frac{r}{r^2 - 2A^2}\left[\frac{1}{2\pi}(q_0 + q(r)) - \frac{A}{c}I_\phi(r)\right]$$

$$B_z = B_0 + \mu_0\left[I_\phi - \left(1 + \frac{A^2}{r^2 - 2A^2}\right)I_\phi(r) - \frac{cA}{r^2 - 2A^2}\frac{q_0 + q(r)}{2\pi}\right] \tag{4}$$

$$B_\phi = \frac{\mu_0}{2\pi r}I_z(r)$$

where $q(r)$, $I_\phi(r)$ and $I_z(r)$ are the net charge per unit length, azimuthal current per unit length, and total current flowing in the z-direction, all contained within a cylindrical shell with outer radius $r$ and inner radius $r_0$. $I_\phi$ is the total azimuthal current per unit length, $I_z$ is the total current flowing in the z-direction, and $B_0$ is the external magnetic field, which for the purposes of this paper we assume is parallel to the z-axis. One immediate implication of eq's 4 is that if A and $I_\phi$ are nonzero, then the charge per unit length of the spinning string must be nonzero. This is reminiscent of the result [4] that a rotating black hole in the presence of an external magnetic field must acquire a charge.

Eq.s (4) describe the electromagnetic fields surrounding a spinning string with fixed external charges and currents. However, in a vacuum where the charges are free to move a fixed configuration of charges and currents will in general not be stable because of magneto-hydrodynamic forces. Substituting the solution (4) into the force-free condition $\bm{j} \times \bm{B} = \rho \bm{E}$ we find that in order for the configuration of fields and charges to be stable the current and charge densities must satisfy the integral equation:

$$j^z \frac{I_z(r)}{2\pi r} - j^\phi \left\{ \frac{B_0}{\mu_0} - I_\phi(r) - \frac{1}{r^2 - 2A^2} \left[ \frac{cA}{2\pi}(q_0 + q(r) - A^2 I_\phi(r)) \right] \right\}$$

$$- \frac{c^2 \rho r}{r^2 - 2A^2} \left[ \frac{q_0 + q(r)}{2\pi} - \frac{A}{c} I_\phi(r) \right] = 0$$

(5)

We expect that in general this equation must be solved numerically. However, since some of the terms in Eq. (5) are also singular at $r=\sqrt{2}A$, it might useful to search for solutions to Eq. (5) where the sum of the singular terms on the l.h.s. of Eq. (5) vanish; i.e.

$$c^2 r \rho(r) \left[ \frac{q_0 + q(r)}{2\pi} - \frac{A}{c} I_\phi(r) \right] = j^\phi(r) \left[ \frac{cA}{2\pi}(q_0 + q(r) - A^2 I_\phi(r)) \right] \quad (6)$$

Remarkably if we assume that the charge and current densities outside the spinning string have the power law behaviors

$$\rho = \rho_0 \left(\frac{r_0}{r}\right)^n, \quad j^\phi = j_0^\phi \left(\frac{r_0}{r}\right)^{n-1}, \quad j^z = j_0^z \left(\frac{r_0}{r}\right)^{n-2}, \quad (7)$$

then Eq. (6) will be satisfied identically if the charge per unit length $q$ outside $r_0$ and total azimuthal current per unit length $I_\phi$ outside $r_0$ satisfy

$$q = -2\pi \frac{A}{c} I_\phi \quad (8)$$

In addition, the non-singular terms on the l.h.s. of Eq.(5) will approximately cancel near to $r_0$ if the total current $I_z$ flowing parallel to the spinning string satisfies:

$$I_z = \pm 2\pi r_0 \left(\frac{n-2}{n-4}\right)^{1/2} \left[ I_\phi \left(\frac{B_0}{\mu_0} + I_\phi\right) \right]^{1/2} \quad (9)$$

There are solutions to this equation for values of $\mu_0 I_\phi$ with either the same sign or the opposite sign as $B_0$. However, solutions where $\mu_0 I_\phi$ has the opposite sign as $B_0$ seem to us more natural since induced currents for free charges in a vacuum are normally diamagnetic. In order that the total current flowing in z-direction $I_z$ be finite the value of n determining the fall-off of the current and charge densities must be > 4. In Fig. 1 we show the force-free polar and azimuthal currents and charge per unit length just outside the spinning string as a function of $r/r_0$ for the value n=5. In plotting these curves it was assumed that $\mu_0 I_\phi = -2B_0$. It can be seen that the currents and total charge reach their asymptotic values fairly rapidly, which implies that the self-generated magnetic fields are closely aligned with the z-axis. In Fig. 2 we show the three dimensional shape of the force-free magnetic field and distribution of associated electrical currents for $r=2A$, where the azimuthal magnetic field is a maximum. Although the self-generated magnetic fields in this example point along the z-axis, the external field far from the string points in the opposite direction.

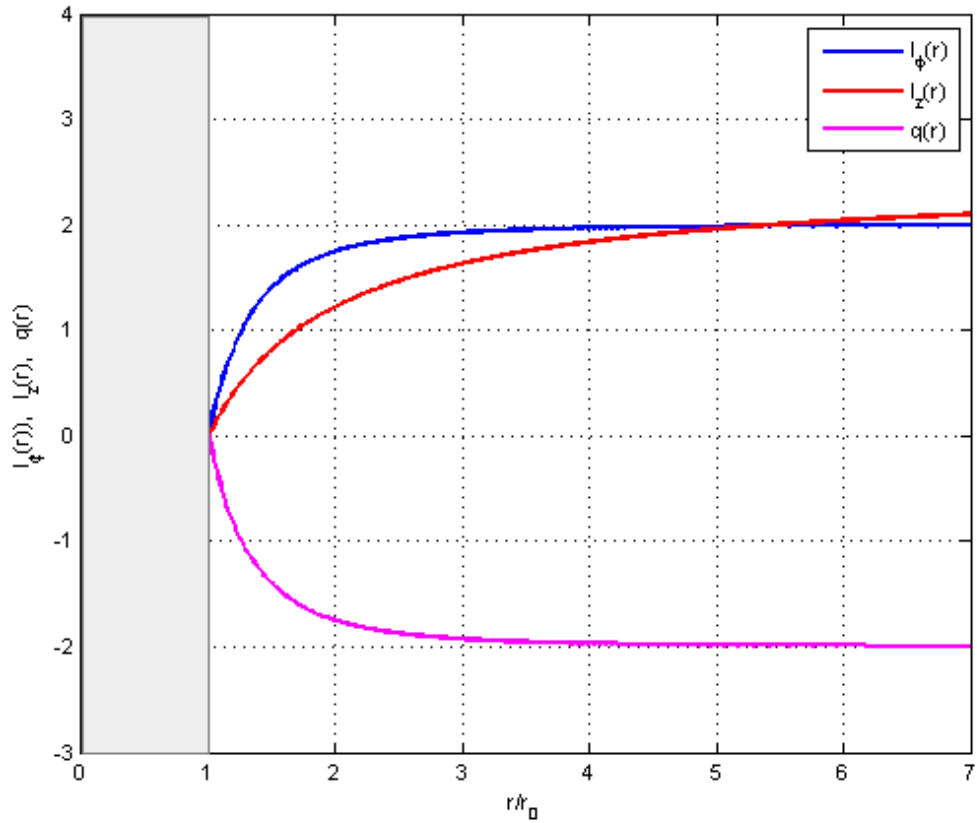

Figure 1. Force free currents $I_\phi(r)$ and $I_z(r)$ and charge per unit length q(r), measured in the natural units $|B_0|/\mu_0$, $2\pi r_0|B_0|/\mu_0$, and $2\pi A|B_0|/c\mu_0$ respectively, as a function of $r/r_0$. These curves assume n=5, $\mu_0 I_\phi = -2B_0$, and $r_0 = \sqrt{2}A$.

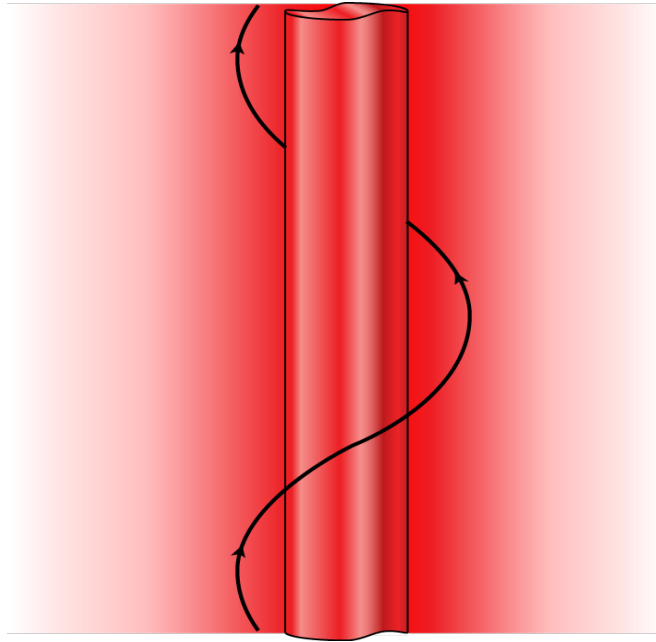

Figure 2. 3-dimensional structure of the force-free magnetic field for $r=2A$ (black line) and associated polar electrical current density (red coloring) surrounding the core of a spinning string.

The magnetic field and currents shown in Fig. 2 are remarkably similar to those inferred from VLBLI mm-wave observations of the relativistic jet that emerges from the center of the active galaxy BL Lac (cf Fig 5 of ref. 2). The success of our simple spinning string magnetosphere model in explaining the initial magnetic structure of this galactic jet is perhaps surprising because the global space-time of a spinning string is quite different from that of a Kerr black hole. Of course, when the angular momentum per unit mass of a Kerr black hole is comparable to its limiting value 2GM/c, one might imagine that the spinning string space-time is simply mimicking the external space-time of a Kerr black hole in its equatorial plane. On the other hand, when the angular momentum per unit mass of the Kerr black hole is much smaller than its limiting value, the external space-time outside the event horizon is quite different from that of a spinning string. Thus it will be particularly interesting to see whether the agreement between our spinning string magnetosphere model and the magnetic structure of jets from activie galaxies persists in cases where the angular momentum per unit mass of the compact object can be shown to be much smaller than the Kerr limiting value 2GM/c. If the success of the spinning string model persists in such cases, then the agreement would suggest that the jets from active galactic nuclei actually originate inside the compact object itself, rather than in the external space-time as in the Blandford-Znajek theory.

Fortunately in the not too distant future the resolution of mm-wave VLBLI observations may improve to the point where it will be possible to directly check whether the jets from active galactic nuclei do indeed originate from within the compact object itself. Such an observation would support the suggestion [5] that the interior space-time of compact objects is quite different from that predicted by general relativity. Indeed it has been proposed [6] that rotating compact objects have a non-singular interior and vortex-like core for which the spinning string space-time might be a reasonable model. One general feature of the jets from compact objects that already hints that this is the case is the extraordinary collimation of the jets emanating from active galaxies. This high degree of collimation is difficult to explain with the Blandford-Znajek theory but, as is evidently the case for our force-free magnetosphere solution (cf. Fig2), would be a natural consequence of a spinning string model for the interior space-time of a massive compact object.

[1]Lawrence Livermore National Laboratory, Livermore, California 94550-808
[2]Advanced Systems Development, NAWC-WC, China Lake, CA 93555
*email: barbierijf@hughes.net

**References**

1. Blandford, R. D. & Znajek, R. L. "Electromagnetic extraction of energy from Kerr black holes", *Mon. Not. R. Astr. Soc.* 179, 433-456 (1977).
2. Marshev, A. P. *et. al*. "The inner jet of an active galactic nucleus as revealed by a radio-to-gamma-ray outburst" *Nature* 452, 966-969 (2008).
3. Mazur, P. O. "Spinning cosmic strings and quantization of energy" *Phys. Rev. Lett.* 57, 929-932 (1986).
4. Wald, R. M. "Black holes in a uniform magnetic field", *Phys. Rev. Lett.* 10, 1680-1685 (1974).
5. Chapline, G., Hohlfeld, E., Laughlin, R. B., & Santiago, D. "Quantum phase transitons and the breakdown of classical general relativity" *Phil Mag* B, 81, 235-254 (2001).
6. Chapline, G., and Marecki, P., arXiv:0809.1115v2 [astro-ph].